\newcommand{\teff}{T$_{\rm eff}$ }

\newcommand{\eqw}{$W_{\lambda}$}

\documentclass{iaus}
\usepackage{graphicx}

\title[\teff Scale of Metal Poor Dwarfs: Li and O Abundances] 
{The Temperature Scale of Metal-Poor Dwarfs: Lithium and Oxygen Abundances
}

\author[Mel\'endez, Shchukina, Ram\'{\i}rez \& Vasiljeva]   
{Jorge Mel\'endez$^1$, 
Nataliya G. Shchukina$^2$, \break
Iv\'an Ram\'{\i}rez$^3$
\and Irina Vasiljeva$^2$}

\affiliation{$^1$Department of Astronomy, Caltech, 1200 E. California Blvd, Pasadena, CA 91125\\
[\affilskip]
$^2$Main Astronomical Observatory, National Academy of Sciences, \\
27 Zabolotnogo Street, Kiev 03680, Ukraine\\[\affilskip]
$^3$Department of Astronomy, University of Texas at Austin, RLM 15.306, TX 78712
}

\pubyear{2005}
\volume{228}  
\pagerange{1--7}
\date{?? and in revised form ??}
\jname{From Lithium to Uranium: Elemental Tracers of Early Cosmic Evolution}
\editors{V. Hill, P. Fran\c{c}ois \& F. Primas, eds.}
\begin{document}

\maketitle

\begin{abstract}
We employ a new Infrared Flux Method (IRFM) temperature scale 
(Ram\'{\i}rez \& Mel\'endez 2005a,b) in order to determine 
Li, O, and Fe NLTE abundances in a sample of relatively unevolved 
(dwarfs, turn-off, subgiants) metal-poor stars. We show that the 
analysis of the permitted OI triplet and FeII lines leads to a plateau 
in [OI/FeII] over the broad metallicity range $-3.2 <$ [Fe/H] $< -$0.7, 
independent of temperature and metallicity, and with a star-to-star 
scatter of only 0.1 dex. The Li abundance in halo stars is also found 
to be independent of temperature and metallicity (Spite plateau), 
with a star-to-star scatter of just 0.06 dex over the metallicity 
range $-3.4 <$ [Fe/H] $< -1$. Our Li abundance (Mel\'endez \& Ram\'{\i}rez 2004) 
is higher than previously reported values, but still lower than the 
primordial abundance suggested by WMAP data and BBN.
\end{abstract}

\firstsection 
\section{New IRFM Temperature Scale}

Recently, Ram\'{\i}rez \& Mel\'endez (2005b, hereafter RM05b) have 
derived a new IRFM \teff scale, which is based
on a homogeneous analysis of more than one thousand stars
for which IRFM temperatures were obtained
employing updated atmospheric parameters
(Ram\'{\i}rez \& Mel\'endez 2005a, hereafter RM05a).
The main improvements compared with previous works 
are a better coverage of the atmospheric parameters
space (\teff, log $g$, [Fe/H]), the use of up-to-date 
metallicities, and the fit of any trend in the residuals (eliminating in
this way spurious trends in the \teff:color:[Fe/H] relations).
The use of updated metallicities and the good
coverage of the atmospheric parameters space
were crucial to derive reliable \teff calibrations, 
greatly helping to distinguish noise from 
real trends with metallicity.

The new IRFM \teff scale (RM05a,b) is in rough agreement
with the scale by Alonso et al.(1996), except for very metal-poor 
F and early G dwarfs, for which we have obtained \teff
calibrations based on a larger sample. According to the
new IRFM \teff scale, the temperature of metal-poor stars close 
to the turn-off is highly dependent on metallicity. Even though
a large dependence of \teff with metallicity in F/G halo dwarfs
has been criticized in the literature (Ryan et al. 1999),
we must remember from stellar evolution that the turn-off of metal-poor stars 
becomes hotter as metallicity decreases 
(see early works by I. Iben, Jr. and collaborators). 
This is a non-negligible effect that accounts for an increase of 600-700 K in
the \teff of the turn-off for a decrease in metallicity from
[Fe/H] = $-1$ to $-3$ (Mel\'endez et al. 2005).

\section{Impact on Lithium Abundances}

Mel\'endez \& Ram\'{\i}rez (2004) have recently
determined the Li abundance in halo stars
employing high quality (\eqw /$\sigma \geq$ 10) literature data,
the new IRFM \teff scale (RM05a,b), and NLTE
corrections by Carlsson et al. (1994).
The Li Spite plateau metal-poor stars (those with \teff $>$ 6000 K)
have Li abundances independent of metallicity and temperature,
with a mean abundance A(Li) = 2.37 and a 
star-to-star scatter of only 0.06 dex over the
broad metallicity range $-3.4 <$ [Fe/H] $< -1$.
The plateau is extremely flat, with essentially zero slope
within the uncertainties (Mel\'endez \& Ram\'{\i}rez 2004). 

On the other hand, Ryan et al. (1999) found a trend of Li abundance
with metallicity, and extrapolating to zero metals they found
A(Li) = 2.0, which is much lower than the primordial Li abundance
derived from WMAP data and Big Bang Nucleosynthesis (A(Li) $\approx$ 2.6).
The very low Li abundance determined by Ryan et al. (1999) is
due to the much lower \teff adopted by them. 
Even though the use of the hotter \teff scale of RM05a,b helps 
to alleviate the difference between the Li stellar abundances 
and the WMAP+BBN prediction, a discrepancy still remains.

\section{Impact on Oxygen Abundances from the OI triplet}

We have applied the new IRFM \teff scale to 31 unevolved halo stars
close to the turn-off, employing recent high S/N literature
data of the OI triplet and FeII lines.

The analysis has been performed in LTE and NLTE, employing
the code NATAJA (Shchukina \& Trujillo Bueno 2001; Shchukina
et al. 2003, 2005). We have used up-to-date atomic data
for the oscillator strengths and damping constants of the
OI triplet and FeII lines. The list of FeII lines has been
updated from the one used by Mel\'endez \&
Barbuy (2002), including the latest laboratory works.

Unlike most previous works in the literature based
on the OI triplet, we have obtained a constant [O/Fe]
$\approx$ +0.5 dex over the broad metallicity range $-3.2 <$ [Fe/H] $< -0.7$,
with a small star-to-star scatter of about 0.1 dex. 
The flat [O/Fe] ratio is essentially
due to the use of the new \teff scale by RM05a,b.
Details will be presented shortly in Mel\'endez et al (2005). 

\begin{acknowledgments}
J.M. thanks partial support from NSF grant AST-0205951 to J. G. Cohen,
and acknowledges the support of the American Astronomical Society 
and the NSF in the form of an International Travel Grant.
I.R. acknowledges support from the Robert A. Welch Foundation
of Houston, Texas, to D. L. Lambert.

\end{acknowledgments}


\end{document}